\documentclass{appolb}
\usepackage{graphicx,amsmath,amssymb}

\begin{document}
\title{Statistical dynamical mean-field description of strongly correlated 
disordered electron-phonon systems
\thanks{Presented at the Strongly Correlated Electron Systems 
Conference, Krak\'ow 2002}%
}
\author{Franz X.~Bronold$^{(1,2)}$ and Holger Fehske$^{(2,3)}$
\address{$^{(1)}$ Institut f\"ur Theoretische Physik, 
Universit\"at Magdeburg,
D-39016 Magdeburg\\
$^{(2)}$ Physikalisches Institut, Universit\"at Bayreuth,
  D-95440 Bayreuth\\ 
$^{(3)}$ Institut f\"ur Physik, 
Universit\"at Greifswald, D-17487 Greifswald
}}
\maketitle
\begin{abstract}Combining the self-consistent theory of localization 
and the dynamical mean-field theory, we present a theoretical
approach capable of describing both self-trapping 
of charge carriers during the process of polaron formation and disorder-induced
Anderson localization. By constructing  random samples for the 
local density of states (LDOS) we analyze the distribution 
function for this quantity and demonstrate 
that the typical rather than the mean LDOS is a natural 
measure to distinguish between itinerant and localized states. 
Significant polaron effects on the mobility edge are found.
\end{abstract}
The question of how the electron-phonon (EP) interaction 
influences the localization
transition caused by disorder~\cite{An58}, i.e. 
by strong impurity-induced spatial fluctuations in
the potential energy, has been addressed by Anderson
about thirty years ago~\cite{An72}. He called attention to 
the particular importance of EP coupling effects in
the vicinity of the so-called "mobility edge", 
separating itinerant (extended) and localized states.
Nevertheless, there is as yet not much theoretical work 
even for the simplest case of a single electron moving in a 
disordered, deformable medium. 

As a first step towards addressing this problem, in Ref.~\cite{BSB01}
the single-particle Holstein model with site-diagonal,
binary-alloy-type disorder was studied within the dynamical
mean field approximation (DMFA)~\cite{GKKR96}. The DMFA, however, cannot 
(fully) discriminate between itinerant and localized states, mainly because
the randomness is treated at the level of the coherent potential
approximation. In order to remedy this shortcoming, recently the 
authors~\cite{BF02} 
adopted the statistical DMFA (statDMFA)~\cite{DK97} 
to the Anderson-Holstein Hamiltonian,
\begin{equation}
{\cal H}=\sum_i \epsilon_i n_i 
-J\sum_{\langle ij\rangle} (c_i^\dagger c_j^{} +{\rm H.c.}) 
-\sqrt{E_p\Omega}\sum_i (b_i + b_i^\dagger) n_i
+ \Omega \sum_i b_i^\dagger b_i\,,
\label{model}
\end{equation}
where $J$ denotes the electron transfer amplitude, 
$\Omega$ is the frequency of the optical phonon, 
$E_p$ is the polaron shift,
and the on-site energies $\{\epsilon_i\}$ are assumed to be
independent random variables with probability 
density $p(\epsilon_i)=(1/\gamma)\theta(\gamma/2-|\epsilon_i|)$. 
The statDMFA is essentially a probabilistic method (in the sense of 
the self-consistent theory of localization~\cite{AAT73}), based on the  
construction of random samples for the physical quantities of interest. 

As a natural measure of the itinerancy of a polaron 
state, we consider the tunneling rate from a given site,
defined - on a Bethe lattice with connectivity 
$K$ ($\tilde{J}=J\sqrt{K}$) -   
as the imaginary part of the hybridization function  
\begin{equation}
\Gamma_i(\omega)=(\pi \tilde{J}^2/K)\sum_{l=1}^K N_l(\omega),\quad \mbox{where}
\quad N_l(\omega)=-(1/\pi){\rm Im} {\cal G}_l(\omega)
\label{Hybrid}
\end{equation}
is the local density of states (LDOS). The LDOS, directly connected
to the local amplitude of the electron wave function, undergoes a 
qualitative change upon localization implying a vanishing 
tunneling rate $\Gamma_i(\omega)$ for a localized state 
at energy $\omega$. The local 
single-particle Green function and the related hybridization function 
are given by
($z=\omega+i\eta$) 
\begin{equation}
{\cal G}_i(z)={1\over z-\epsilon_i-H_i(z)-{\Sigma}_i(z)}\quad\mbox{and} \quad  
H_i={{\tilde{J}^2}\over K}\sum_{l=1}^{K+1}{1\over{z-\epsilon_l-\bar{H}_l^i-\bar{\Sigma}_l^i}},
\label{GF}
\end{equation}
respectively. We now ignore that the functions on 
the rhs of $H_i$ should be calculated
for the Bethe lattice with the site $i$ removed, i.e. we make the replacement 
$\{\bar{\cal G}_l^i,\;\bar{H}_l^i,\;\bar{\Sigma}_l^i\}\; \leadsto 
\;\{{\cal G}_l,\;H_l,\;\Sigma_l\}$,
and furthermore 
take $K$ as the typical number of terms even for the central site. 
Finally, the EP self-energy contribution
is determined in the limit $K\to \infty$. The self-energy is then 
local and, in terms of a continuous fraction expansion,   
takes the form 
\begin{equation}
\Sigma_l(z) =\frac{ E_p 1 \Omega}{[F_l^{(1)}(z)]^{-1}- \frac{E_p 2\Omega}{[F_l^{(2)}(z)]^{-1}-...}}, 
\label{Sigma}
\end{equation}
with $[F_l^{(p)}(z)]^{-1}=z-p\Omega-\epsilon_i-H_{l}^{(p)}(z)$ and
$H_l^{(p)}(z)=H_l(z-p\Omega$).  Here the energy shift
keeps track of the number of virtual phonons \mbox{($0\!<\!p\!<\!M$)}.
Regardless of the local EP self-energy, the statDMFA  
takes spatial fluctuations of, e.g., the 
LDOS into account and provides an adequate 
description of disorder effects.  Due to the randomness in 
the on-site energies, the tunneling rate 
and consequently the LDOS is a random variable, 
and the question of whether it vanishes or not depends on 
the probability density exhibiting different features 
for itinerant and localized states~\cite{An58,AAT73}.
In particular, the difference between the 
mean and typical LDOS, 
\begin{equation}
{N}^{\rm mean}(\omega)= \frac{1}{N}\sum_i^{N}N_i(\omega)\quad \mbox{and}\quad
N^{\rm typ}(\omega)=\exp 
\left[\frac{1}{N}\sum_i^{N}\log N_i(\omega)\right]
\label{LDOS}
\end{equation}
obtained by the arithmetic and geometric mean of the LDOS, respectively,
is a useful measure to discriminate between extended and localized states.  
${N}^{\rm mean}(\omega)>0$ but $N^{\rm typ}(\omega)=0$ indicates a 
localized state at energy $\omega$.

In the numerical work, we calculated the LDOS by solving a recursion scheme
for $H_{l}^{(p)}$
which depends on $K \varepsilon_j^{\;\prime}\mbox{s}$, 
$K H_{j}^{(p)\;\prime}\mbox{s}$, $\ldots$, and
$K H_{j}^{(p_{max})\;\prime}\mbox{s}$.
Starting from an initial random configuration for the independent 
variables~ 
$H_{l}^{(p)}$, which is successively updated with a sampling technique similar 
to the one described in Ref.~\cite{AAT73},  we constructed 
self-consistent random samples for $H_{l}^{(p)}$, 
using  $K=2$, $N=100\,000$, $M=35$, and $\eta=10^{-8}$.  

Without disorder, the physical properties of 
the Holstein model are determined by two interaction parameters, 
$\tilde{\lambda}=E_p/2\tilde{J}$ 
and $g^2=E_p/\Omega$, and the adiabaticity ratio  
$\tilde{\alpha}=\Omega/\tilde{J}$. Polaron formation sets in
provided that $\tilde{\lambda}\gtrsim 1/\sqrt{K}$ and $g^2 \gtrsim 1$. 
Of course, the internal structure of the polaron depends on 
$\tilde{\alpha}$.
\begin{figure}[!b]
\begin{minipage}{0.4\linewidth}
\includegraphics[width=1.0\textwidth,angle=-90]{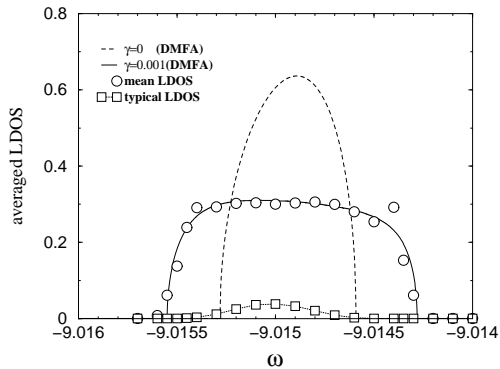}
\end{minipage}\hspace*{1.8cm}\begin{minipage}{0.43\linewidth}
\caption{Mean and typical LDOS for the lowest polaron subband 
in the non-adiabatic strong EP coupling
region ($\tilde{\lambda}=9.0$, $\tilde{\alpha}=2.25$, and $J=0.5$).
The pronounced disorder-induced broadening of the LDOS occurs 
because the variation of the on-site energies $\gamma=0.001$ is on the order 
of the strongly renormalized band width $\tilde{W}$.}
\end{minipage}
\end{figure}
Disorder affects polaron states quite differently 
in the adiabatic ($\tilde{\alpha} \ll 1$), 
non-adiabatic ($\tilde{\alpha} \sim 1$), and antiadiabatic 
($\tilde{\alpha} \gg 1$) cases.
Without EP coupling, i.e. in the pure Anderson model, the critical 
disorder strength needed to localize all states is  
$(\gamma_c/\tilde{W}_0)_{\rm complete}\approx 2.25$, where 
$\tilde{W}_0=4\tilde{J}$. In the weak EP coupling regime,
it has been shown that the quantum interference 
needed for localization is significantly suppressed by 
inelastic polaron-phonon scattering processes~\cite{BF02}:
States above the optical phonon emission threshold
are more difficult to localize than the corresponding bare electron states. 
In the very strong EP coupling regime, extremely weak disorder turns
itinerant  into localized polaron states. Surprisingly,
the ratio $(\gamma_c/\tilde{W})_{\rm complete}$, where 
$\tilde{W}$ is the band width of the lowest polaron subband,
is almost the same as for a bare electron. In fact, in the 
non-adiabatic strong EP coupling 
regime, where the band collapse changes only the overall energy scale, 
disorder affects a polaron in a similar way as a bare electron.
For example, the LDOS and mobility edges are symmetric (cf. Fig.~1). 
\begin{figure}[!t]
\begin{center}
\includegraphics[width=0.4\textwidth,angle=-90]{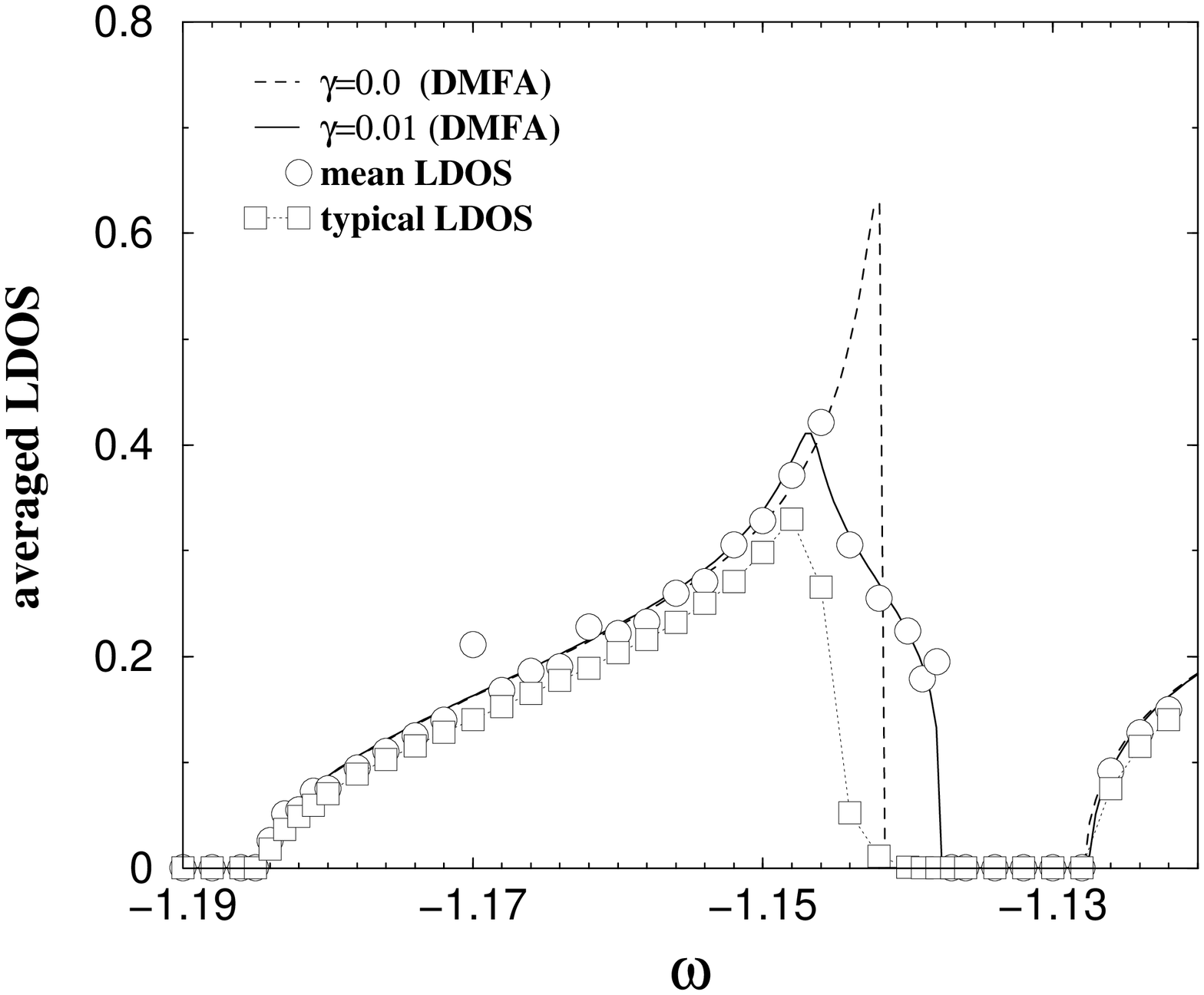}\hspace*{0.5cm}
\includegraphics[width=0.4\textwidth,angle=-90]{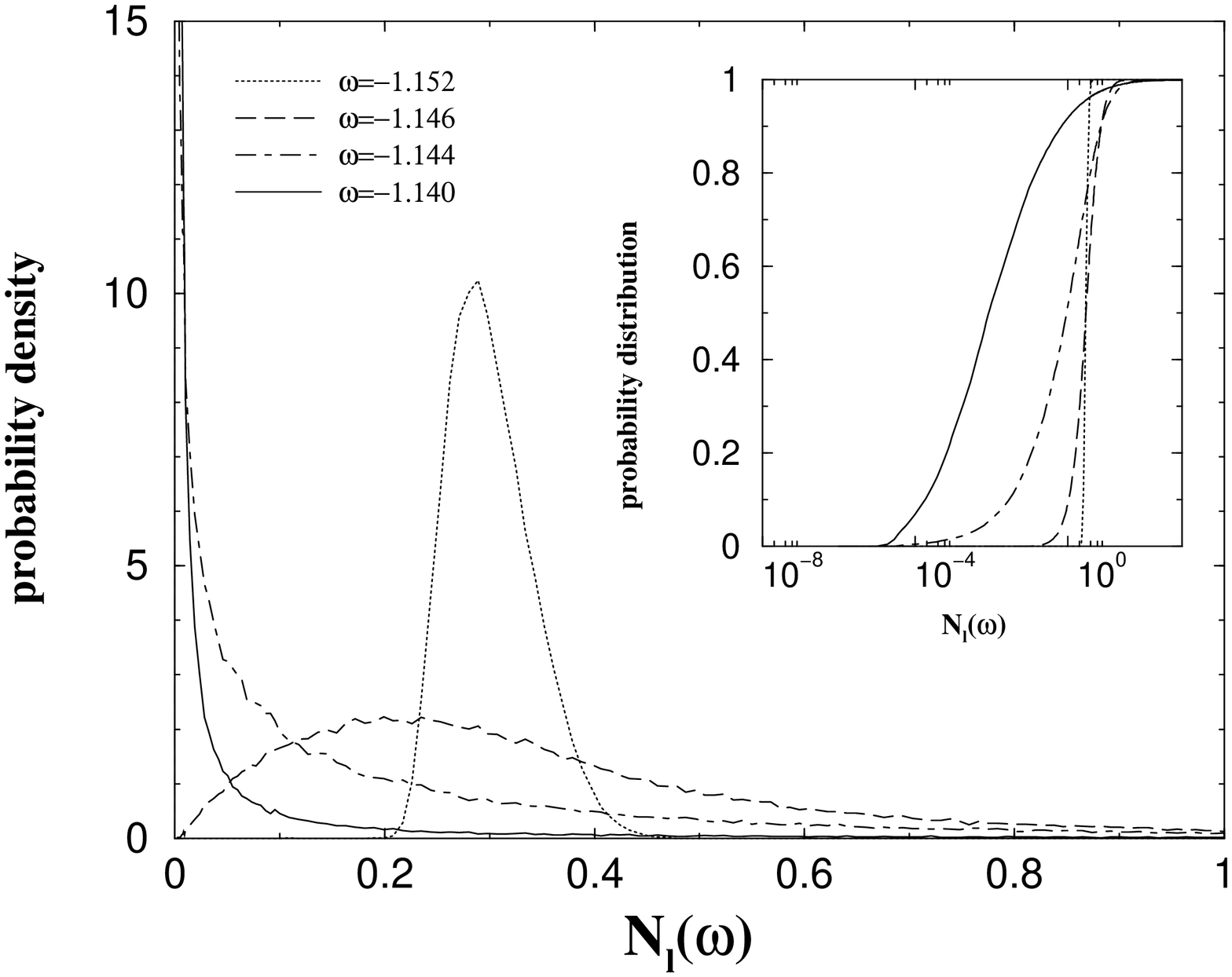}
\end{center}
\caption{Left panel: Mean and typical LDOS 
in the adiabatic intermediate-to-strong EP coupling
region ($\tilde{\lambda}=1.0$, $\tilde{\alpha}=0.25$, $J=0.5$).
Note that $N^{\rm mean}(\omega)$ is almost perfectly approximated
by the DMFA. At about $\omega =- 1.13$ the second polaronic subband 
starts. Right panel: Probability density of the LDOS for four
representative energies $\omega$. 
The inset shows the probability distribution, i.e., the
cumulant of the probability density.} 
\label{fig2}
\end{figure}

In the adiabatic intermediate-to-strong EP coupling
regime the physics is much more involved. Here the band 
dispersion of the lowest subband significantly deviates 
from a rescaled bare band~\cite{FLW97}, leading to a
strong asymmetric LDOS. Specifically, the states at the bottom 
of the subband are mostly electronic and rather mobile due to 
long-range tunneling induced by EP coupling, whereas the states 
at the top of the subband are rather phononic and immobile~\cite{FLW97}. 
As a direct consequence, the states at the zone boundary are 
very susceptible to disorder, i.e. the critical disorder strength 
needed to localize these states is much smaller than for states 
at the bottom, and, from the results for the typical LDOS,
we find asymmetric mobility edges (see Fig.~2). Moreover,
$(\gamma_c/\tilde{W})_{\rm complete}\approx 2.8$, which 
is larger than the corresponding ratio for a bare electron. Thus,   
contrary to naive expectations, at intermediate EP couplings, 
an adiabatic polaron is even more difficult to localize than a 
bare electron. 

It is very instructive to discuss the behaviour of the probability density  
of the LDOS  and the corresponding probability distribution.  
Note that both quantities have to be calculated self-consistently 
within our sampling procedure. The right panel of Fig.~2 shows 
the dramatic change of the probability density of $N_l(\omega)$ when 
the system undergoes the localization transition 
by crossing the mobility edge. In the region of localized states, 
the probability density for the LDOS is broad and very asymmetric and, 
as a consequence, the 
mean LDOS is not representative.   

In conclusion, in terms of the Anderson Holstein model, 
we have demonstrated that the statDMFA, which according to the 
spirit of Anderson's early work~\cite{An58} focuses on distribution functions
and associates typical rather than mean values to physical quantities, 
yields a proper description of disordered electron-phonon systems.


\begin{thebibliography}{}

\bibitem{An58}
P. W. Anderson, Phys. Rev. {\bf 109}, 1498 (1958).


\bibitem{An72}
P. W. Anderson, Nature Phys. Science {\bf 235}, 163 (1972)


\bibitem{BSB01}
F.X. Bronold, A. Saxena, and A. R. Bishop, 
Phys. Rev. B {\bf 63}, 235109 (2001). 

\bibitem{GKKR96}
A. Georges {\it et al.}, 
Rev. Mod. Phys. {\bf 68}, 13 (1996).


\bibitem{BF02}
F.X. Bronold and H. Fehske, 
Phys. Rev. B, accepted for publication  (2002). 

\bibitem{DK97}
V. Dobrosavljevi\'c and G. Kotliar, Phys. Rev. Lett. {\bf 78},
3943 (1997).

\bibitem{AAT73}
R. Abou-Chacra, P. W. Anderson, and D. Thouless,  
J. Phys. C {\bf 6}, 1734 (1973). 


\bibitem{FLW97}
H. Fehske, J. Loos, and G. Wellein, 
Z. Phys. B {\bf 104}, 616 (1997).

\end{thebibliography}
\end{document}